\documentclass[%
 reprint,
 amsmath,amssymb,
 aps,
superscriptaddress]{revtex4-2}

\usepackage{graphicx}
\usepackage{dcolumn}
\usepackage{bm}
\let\PhysLett\pl
\let\pl\relax
\usepackage{unitsdef}
\let\pl\PhysLett

\usepackage{xcolor}
\usepackage{amsmath}
\usepackage[english]{babel}
\makeatletter
\def\l@en{\l@english}
\makeatother

\makeatletter
\def\l@eng{\l@english}
\makeatother

\begin{document}

\preprint{APS/123-QED}

\title{A superinductor in a deep sub-micron integrated circuit}

\author{Thomas H. Swift}
\affiliation{Quantum Motion, 9 Sterling Way, London, N7 9HJ, United Kingdom}
\affiliation{London Centre for Nanotechnology, UCL, London, WC1H 0AH, United Kingdom}
\author{Fabio Olivieri}
\affiliation{Quantum Motion, 9 Sterling Way, London, N7 9HJ, United Kingdom}
\author{Gorka Aizpurua-Iraola}
\affiliation{Quantum Motion, 9 Sterling Way, London, N7 9HJ, United Kingdom}
\affiliation{CIC nanoGUNE Consolider, Tolosa Hiribidea 76, E-20018 Donostia-San Sebastian, Spain}
\author{James Kirkman}
\affiliation{Quantum Motion, 9 Sterling Way, London, N7 9HJ, United Kingdom}
\author{Grayson M. Noah}
\affiliation{Quantum Motion, 9 Sterling Way, London, N7 9HJ, United Kingdom}
\author{Mathieu de Kruijf}
\affiliation{Quantum Motion, 9 Sterling Way, London, N7 9HJ, United Kingdom}
\affiliation{London Centre for Nanotechnology, UCL, London, WC1H 0AH, United Kingdom}
\author{Felix-Ekkehard von Horstig}
\affiliation{Quantum Motion, 9 Sterling Way, London, N7 9HJ, United Kingdom}
\affiliation{Department of Materials Sciences and Metallurgy, University of Cambridge, Charles Babbage Rd, Cambridge
CB3 0FS, United Kingdom}
\author{Alberto~Gomez-Saiz}
\email{alberto@quantummotion.tech}
\affiliation{Quantum Motion, 9 Sterling Way, London, N7 9HJ, United Kingdom}
\affiliation{Department of Electrical and Electronic Engineering, Imperial College London, London SW7 2AZ, United Kingdom}
\author{John~J.~L.~Morton}
\email{john@quantummotion.tech}
\affiliation{Quantum Motion, 9 Sterling Way, London, N7 9HJ, United Kingdom}
\affiliation{London Centre for Nanotechnology, UCL, London, WC1H 0AH, United Kingdom}
\author{M. Fernando Gonzalez-Zalba}
\email{fernando@quantummotion.tech}
\affiliation{Quantum Motion, 9 Sterling Way, London, N7 9HJ, United Kingdom}
\affiliation{CIC nanoGUNE Consolider, Tolosa Hiribidea 76, E-20018 Donostia-San Sebastian, Spain}
\affiliation{IKERBASQUE, Basque Foundation for Science, E-48011 Bilbao, Spain}

\date{\today}

\begin{abstract}
Superinductors are circuit elements characterised by an intrinsic impedance in excess of the superconducting resistance quantum ($R_\text{Q}\approx6.45~$k$\Omega$), with applications from metrology~\cite{Flowers2004} and sensing~\cite{Vine2023} to quantum computing~\cite{bell_quantum_2012, Masluk2012, Grunhaupt2019}. However, they are typically obtained using exotic materials with high density inductance such as Josephson junctions~\cite{bell_quantum_2012, Nguyen2019}, superconducting nanowires~\cite{niepce_high_2019, Srivastava2025} or twisted two-dimensional materials~\cite{Tanaka2025, Banerjee2025, Jha2025}. Here, we present a superinductor realised within a silicon integrated circuit (IC), exploiting the high kinetic inductance ($\sim 1$~nH/$\square$) of TiN thin films native to the manufacturing process (22-nm FDSOI). By interfacing the superinductor to a silicon quantum dot formed within the same IC, we demonstrate a radio-frequency single-electron transistor (rfSET), the most widely used sensor in semiconductor-based quantum computers. The integrated nature of the rfSET reduces its parasitics which, together with the high impedance, yields a sensitivity improvement of more than two orders of magnitude over the state-of-the-art, combined with a 10,000-fold area reduction. Beyond providing the basis for dense arrays of integrated and high-performance qubit sensors, the realization of high-kinetic-inductance superconducting devices integrated within modern silicon ICs opens many opportunities, including kinetic-inductance detector arrays for astronomy~\cite{Day2003} and the study of metamaterials and quantum simulators based on 1D and 2D resonator arrays~\cite{Zhang2023}. 

\end{abstract}

\maketitle 

The discovery of superconductivity and the development of the silicon Integrated Circuit (IC) are two of the most significant developments in physics of the past century. Superconductors led to the appreciation of strong electron correlations which remain a vibrant topic within condensed matter physics and underpin technologies ranging from magnetic resonance imaging, to quantum limited amplifiers, quantum computers and high-sensitivity photon detectors. Meanwhile, modern silicon complementary metal-oxide-semiconductor (CMOS) ICs have become the most complex devices mankind has produced, enabling billions of discrete components with feature sizes below 40 nm to be integrated with high yield and high volume. 

Throughout the development of CMOS IC technology, the materials in the transistor gate stacks have been regularly refined to optimise properties such as dielectric permittivity and workfunction. Meanwhile, the need for cryogenic electronics for quantum computers~\cite{anders2023} and space applications~\cite{hong2008} has motivated the study and operation of CMOS ICs at deep cryogenic temperatures ($\leq10$~K). Models are being developed to support the design of CMOS ICs to operate at such temperatures, capturing the temperature-dependent variations in parameters such as subthreshold swing, carrier mobility, and threshold voltage~\cite{Patra2018}. However, operating silicon ICs at cryogenic temperatures also enables the emergence of profoundly new capabilities arising, for example, from quantum dots as highly non-linear circuit elements~\cite{Oakes2022} and the onset of superconductivity~\cite{Noah2024} in the materials used within the gate stack and other metal layers. In addition to the low losses characteristic of materials in their superconducting state, the inertia of the Cooper pairs leads to an additional ‘kinetic’ contribution to inductance which may be used to produce more compact inductors or form the basis of protected qubit designs and, quantum-limited amplifiers and detectors~\cite{Grunhaupt2019}.

In this Article, we present the realisation of superconducting circuit elements within a 22~nm CMOS process, exploiting thin-film TiN present within the technology gate stack~\cite{scheiblin2017}. We characterise the kinetic inductance of the thin superconducting films found in two variants of poly-resistor available in the 22-nm FDSOI process from GlobalFoundries. We use these films to create integrated TiN superinductors over four orders of magnitude more compact than conventional spiral inductors used in CMOS circuits. This result provides the basis for a monolithic integration of digital and analogue electronics used in signal processing and quantum error decoding~\cite{Bausch2024} together with compact superconducting readout elements of silicon-based qubits, providing significant advantages in scalability and reduced latency towards an integrated quantum computing system~\cite{Gonzalez-Zalba2021}. In particular, we demonstrate an integrated rfSET, the most widely used sensor for semiconductors spin qubits~\cite{Schoelkopf1998}, and show improvement of over two-orders of magnitude in sensitivity with respect to the state-of-the-art. We envision that integrating compact superconducting elements with the vast capabilities of modern silicon ICs may open up many further opportunities for innovation in quantum computing as well as low-power cryoelectronics.

\begin{figure}[ht!]
\includegraphics[width=\columnwidth]{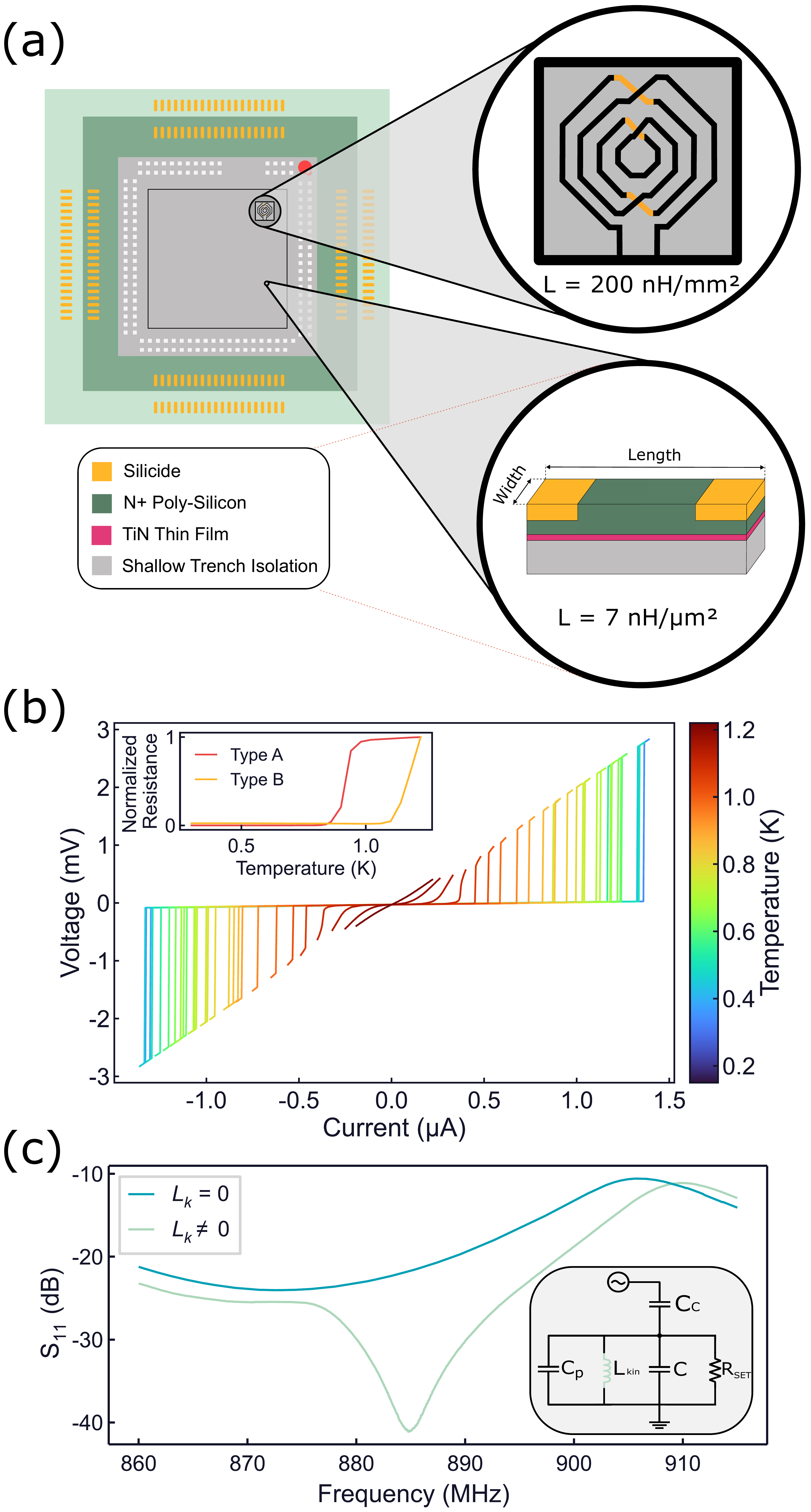}
\caption{\label{fig:1}\textbf{A titanium nitride CMOS superinductor.} (a) 22-nm CMOS chip diagram with insets showing sizes of the spiral inductor and TiN thin film superinductor. (b) Four point I-V curve for different temperatures for Type B superconducting TiN thin film, with the inset showing the resistance vs temperature for both Type A and Type B thin film types, normalized to the maximum resistance of Type A. (c) S$_{11}$ measurement of the resonant circuit shown in the inset, for the superconducting (green) and normal (blue) states of the film, the latter achieved using an rf power above the critical rf power.}
\end{figure}

\section{\label{sec:1} Parametric Characterization of the Superinductors}
Spiral inductors used in CMOS circuits have a typical inductance of $50-200$~nH/mm$^{2}$~\cite{wang_146196-db-nf_2023} and as a result consume a significant fraction of the chip area, as illustrated in Fig.~\ref{fig:1}(a). Here, we study inductors based on two variants of poly-resistors (also shown in Fig.~\ref{fig:1}(a)) possessing a TiN layer which can become superconducting at low temperature. We label the device variants Type A and Type B, with the main difference being the use of different doping profiles. To determine the properties of the superconducting thin films, we perform four-point measurements of two types of TiN-based structures (types A and B) of varying width ($W$ = 0.36 - 3.6~$\mathrm{\mu}$m) and length ($L$ = 0.36 - 50~$\mathrm{\mu}$m). We place the films in an array (referred to as a farm) and address them sequentially with an on-chip multiplexer~\cite{thomas_rapid_2025}. We implement the farm on a 22-nm FDSOI IC powered by a set of 0.8~V and 1.8~V supplies, leading to a small static power dissipation that sets the chip base temperature to 350~mK. We first measure temperature-dependent I-V curves as shown in Fig.~\ref{fig:1}(b). We identify the switching current ($I_{\mathrm{sw}}$) as the point at which TiN transitions from the superconducting to the normal state. For the dimensions studied, and assuming a film thickness $t=6$~nm, we find critical current densities, $J_c =94 \pm 27$ A/mm$^2$ and 260 $\pm$ 63 A/mm$^2$ for Type A and Type B thin films, respectively. We also extract the critical temperature from the resistance vs. temperature plot in the inset of Fig.~\ref{fig:1}(b), and obtain $T_{\mathrm{C}}=0.75$~K and 1.1~K for Type A and B, respectively.\\

To characterise the kinetic inductance ($L_{\mathrm{k}}$), we embed the superconductor in an integrated resonant circuit schematically depicted in Fig.~\ref{fig:1}(c). 
We complete the resonant circuit with metal-oxide-metal (MOM) capacitors for line coupling ($C_{c}$ = 164~fF) and for the resonator ($C$ = 114 fF). In the model, we include the intrinsic capacitance of the inductor ($C_p=6.3$~fF) which appears in parallel (see Section~\ref{sec:Methods:Res_Design}). Additionally, we add a transistor whose channel resistance ($R_\text{SET}$) appears in parallel with the inductor. The variable resistance of the transistor allows controlling the dissipation in the resonator and hence its coupling to the drive line. In addition, the small gate length ($L_g=28$~nm) and narrow channel width ($w=80$~ nm) allow the formation of a quantum dot in the silicon channel~\cite{thomas_rapid_2025}, as we shall see in Section \ref{sec:2}. To demonstrate the resonant behaviour of the circuit, we plot the reflection coefficient versus frequency in Fig.~\ref{fig:1}(c). In the superconducting state (green trace), we observe reduced reflection at the resonant frequency of the circuit ($f_0=884$~MHz), whereas such a resonance feature is absent in the normal state (blue trace). These results demonstrate the presence of an inductive element arising from the superconductivity of the TiN thin film. We extract the value of the inductance by using $L_{\mathrm{K}} = 1/[(2\pi f)^2C_{\mathrm{tot}}]$ where $C_{\mathrm{tot}}= C_{\mathrm{c}}+C+C_p$ giving an estimated $L_{\mathrm{K}}$ = 149(131)~nH and kinetic inductance per square of $L_{\mathrm{K}}$ = 1.07(0.94)~nH/$\square$ for type A(B).\\

\begin{figure}[ht!]
\includegraphics[width=\columnwidth]{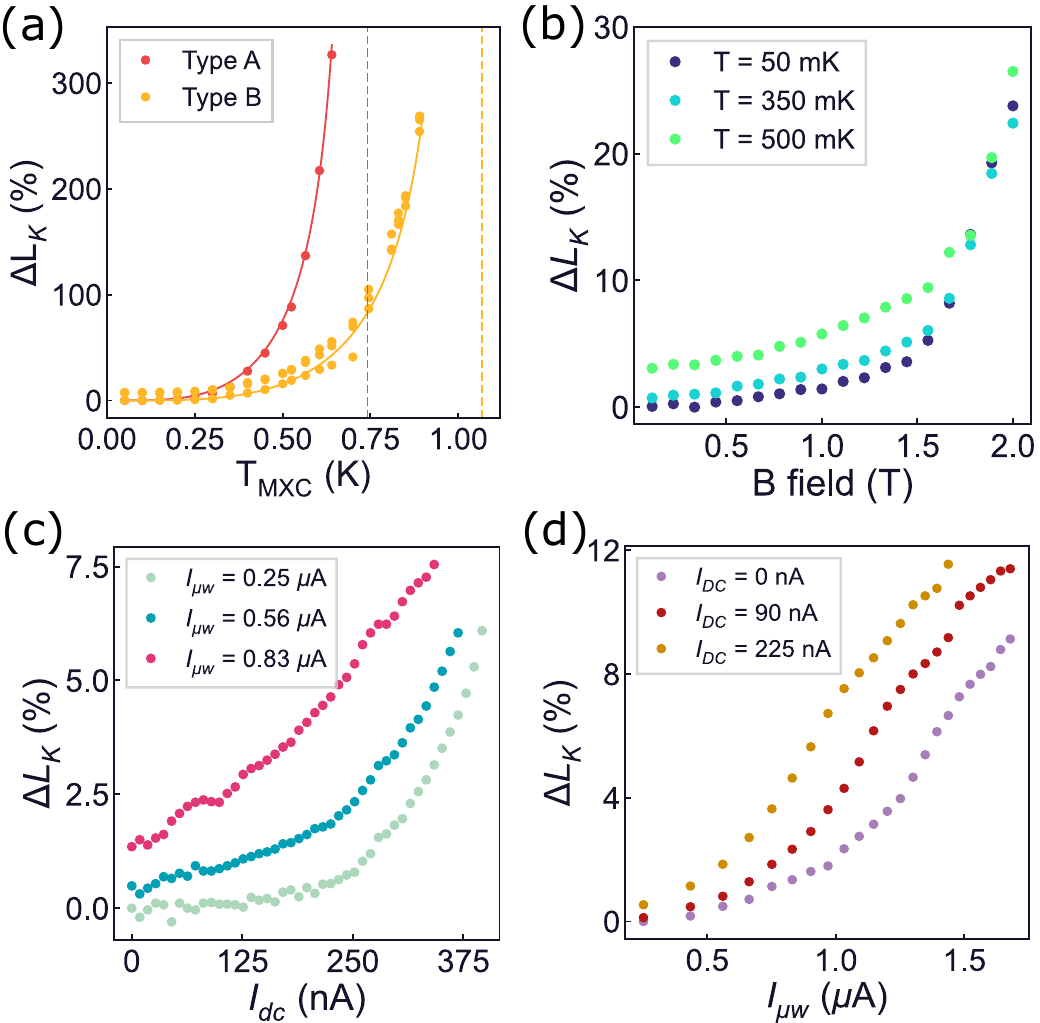}\caption{\label{fig:2} \textbf{Parametric characterisation.} (a) Dependence of kinetic inductance on temperature for TiN thin films of Type A (red, one device) and Type B (orange, three devices) with a fit to Eq.~1.  For Type A: (b) magnetic field dependence of kinetic inductance for different mixing chamber temperatures and (c) \& (d) kinetic inductance as a function of dc current (rf power) for different rf powers (dc currents).}
\end{figure}

In addition to the reduced size of the inductor resulting from the use of kinetic inductance, there is also the added benefit of inductance tunability. In Fig.~\ref{fig:2}, we explore its dependence with temperature, magnetic field and dc and rf currents. First, in Fig.~\ref{fig:2}(a), we show the change in $L_{\mathrm{K}}$ as a function of the mixing chamber temperature of the dilution refrigerator ($T_\text{MXC}$) for the Type A (red) and Type B (orange) thin films. We see the inductance plateauing at low temperatures and diverging near the critical temperature. We model the temperature dependence of the kinetic inductance considering the Cooper pair density ($n_{\mathrm{s}}$) in the Ginzburg-Landau theory as $n_{\mathrm{s}}\left(T\right) \approx n_{\mathrm{s}}\left(0\right)\left(1-T/T_{\mathrm{c}}\right)$ and find, 

\begin{equation}
    L_{\mathrm{K}} = \frac{m}{2n_se^2}\frac{L}{Wt} \approx \frac{L_{\mathrm{K}}\left(0\right)}{1-\frac{T}{T_{\mathrm{C}}}},
\label{eq:Lk_T_dependence}
\end{equation}

\noindent where $m$ is the mass of the electron and $e$ the elementary charge~\cite{annunziata_tunable_2010,tinkham_introduction_2015}. We fit Eq.\eqref{eq:Lk_T_dependence} to the data considering an effective temperature $T=\sqrt[4]{T_{\mathrm{MXC}}^{4}+T_\text{el}^{4}}$ that accounts for the minimum chip temperature ($T_\text{E}$) associated with the static on-chip power dissipation~\cite{Noah2024}. Here, the fourth root indicates thermalisation through insulating materials. We find $T_{\mathrm{el}}$ = 350~mK and the dashed vertical lines indicate the fitted $T_{\mathrm{C}}$ which aligns with the values measured using dc techniques as shown in Fig.~\ref{fig:1}(b). Using the maximum value of measured inductance (807~nH) and the parasitic capacitance of the inductor ($C_{par}^{L}$ = 6.3~fF), we determine the maximum impedance of the inductor as $Z_{L}=\sqrt{L_K/C_{par}^{L}}$ = 11.5~k$\mathrm{\Omega}$ which well exceeds $R_Q$.
\\

Next, in panel b, we plot the effect on $L_{\mathrm{K}}$ of a magnetic field applied in the plane of the film for several temperatures. As expected, the value of $L_{\mathrm{K}}$ increases with field as the Cooper pair density decreases. The TiN films show resilience to in-plane fields up to 2~T making them compatible with spin qubits. In addition, in panels c and d, we show how the kinetic inductance increases with the dc and rf currents as given by, \cite{clem_geometry-dependent_2011,frasca_three-wave-mixing_2024},


\begin{equation}
    L_{\mathrm{K}} = L_{\mathrm{K,0}}\left(1+\frac{I_{dc}^2}{I_{*}^2}+\frac{2I_{dc}I_{rf}}{I_*^2}+\frac{I_{rf}^2}{I_{*}^2}\right).
\label{eq:Lk_current}
\end{equation}


We apply the dc current by varying the gate and source-drain voltages of the transistor and the rf current by changing the applied rf power. We find a quadratic dependence for low rf powers which becomes more linear for higher powers as the second term in Eq.\ref{eq:Lk_current} begins to dominate. Fig.~\ref{fig:2}(d) then highlights the symmetry of Eq.\ref{eq:Lk_current} showing the dependence of $L_{\mathrm{K}}$ on rf power for different dc currents. This change in $L_{\mathrm{K}}$ as a function of rf and dc inputs represents a non-linearity that allows the possibility of achieving parametric amplification through three-wave or four-wave mixing processes~\cite{Parker2022}. For the case of $I_{\mathrm{dc}} = 0$, the non-linearity of the kinetic inductance can be directly compared to a self-Kerr characterized by a coefficient of $12.0\pm 0.2$ and $5.29\pm 0.06$~Hz/photon for Type A and B, respectively.

\section{\label{sec:2}Use Case: A radio-frequency single-electron transistor}

\begin{figure*}[ht!]
\includegraphics[width=\textwidth]{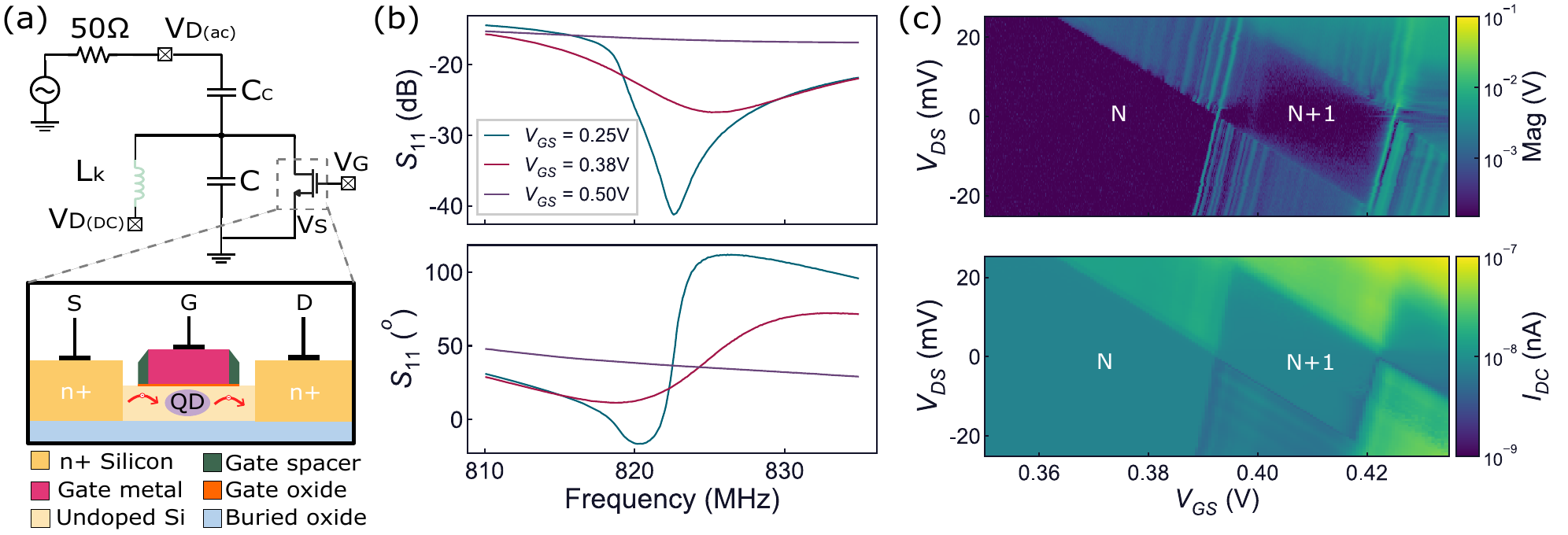}
\caption{\label{fig:3} \textbf{Integrated rfSET.} (a) Circuit schematic including the rf source, coupling capacitor $C_c$, kinetic inductor $L_K$, equivalent resonator capacitance $C$ and field-effect transistor used to form the SET. Inset) Schematic cross-section of the SET showing a charged island at the Si/SiO$_2$ interface directly under the gate as well as the flow of the single-electron AC current (red arrows). (b) $S_{11}$ magnitude and phase of the resonant circuit as a function of the gate voltage of the SET demonstrating the sensitivity to changes in the SET resistance. (c) Magnitude of the reflected rf signal (top) and drain
current (bottom) as a function of the gate-source and drain-source voltage, showing regions of charge stability, i.e. Coulomb diamonds. The labels indicate the charge configuration where $N$ indicates a discrete charge offset. The positive slope features inside the conductive regions are associated to 1d density of states in the SD contacts.}
\end{figure*}

To exemplify the benefits of the compact integrated inductor, we present a radio-frequency single-electron transistor (rfSET), the most widely used sensor for spin qubits in semiconductor-based quantum computers. The SET is a three-terminal Coulomb blockade device made up of a conductive island capacitively coupled to a gate electrode (G) and tunnel-coupled to two electronic reservoirs (source-S and drain-D), see Fig.~\ref{fig:3}(a). The source-drain resistance of the SET, $R_\text{SET}$, is highly sensitive to its electrostatic environment and hence can be used as a sensitive charge sensor. When combined with spin-to-charge conversion techniques~\cite{Ono2002, elzerman_single-shot_2004}, the SET can determine the spin state of semiconductor spin qubits from a single charge tunneling event. To enhance its bandwidth and allow faster measurements, the SET is typically embedded in an $LC$ matching network that converts the high source-drain impedance of the device ($R_\text{SET}\geq 8R_Q$) to the 50~$\Omega$ of a transmission line (see Fig.~\ref{fig:3}(a)), allowing low-noise cryogenic amplification in transmission or reflection~\cite{Vigneau2023}. The rfSET is the most sensitive charge sensor, but the size of the inductor, orders of magnitude larger than that of typical spin qubits, has compromised the scalability of spin-based quantum computers. Here, we use the TiN superinductor to address this challenge whilst also enhancing the sensitivity of the sensor. 

In our experiment, we form the SET in a narrow channel ($w=80$~nm), short gate length transistor ($L_g=28$~nm), see panel a. A conductive island forms in the intrinsic silicon channel when we apply a gate-source voltage ($V_\text{GS}$) near the threshold of the transistor. Tunnel barriers form in the ungated and low-doped silicon region directly under the gate spacers (dark green trapezoids). The reflection coefficient, $S_{11}$, of the $LC$ resonator is dependent on the impedance of the SET as shown in Fig.~\ref{fig:3}(b) where we observe changes in the magnitude and phase around the resonance frequency ($f_r=823$~MHz). We use the sensitivity to impedance changes to probe the charge state of the SET by applying a gate-source ($V_{\mathrm{GS}}$) and drain-source voltage ($V_{\mathrm{DS}}$) as shown in Fig.~\ref{fig:3}(c). The top panel shows the magnitude of the reflected signal, whereas the bottom panel is the measured current through the SET. Both panels show the signature of discrete charging, i.e. diamond shapes, associated with the formation of the conductive island.

\section{\label{sec:3} SNR \& Minimum Integration Time}

\begin{figure}[ht!]
\includegraphics[width=\columnwidth]{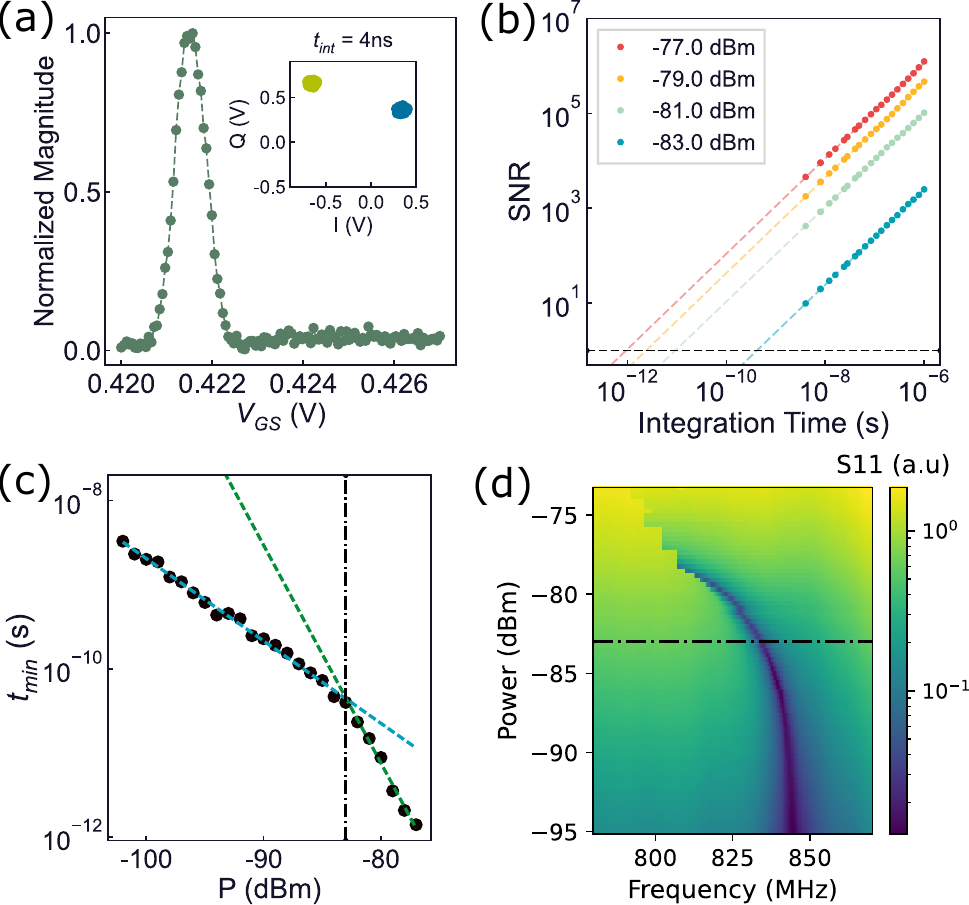}
\caption{\label{fig:4} \textbf{Sensitivity benchmark.} (a) Plot of the magnitude of the reflected signal versus gate-source voltage for the Coulomb peak used in the sensitivity benchmark. Inset) Plot in the IQ plane of 1~$\mu$s time traces at the top (green) and bottom (blue) of the Coulomb peak using an integration time of 4~ns. (b) SNR vs integration time for different rf powers with linear fits extrapolating to SNR = 1. (c) Minimum integration time as a function of applied rf power with fits to the low power (blue) and high power (green) shown by the dashed lines. (d) $S_{11}$ measurement of the resonant circuit versus frequency and power showing a resonance frequency reduction as the power is increased due to the non-linear behaviour of the TiN inductance. At even higher powers, the resonator becomes normal.}
\end{figure}

To benchmark the charge sensitivity of the integrated rfSET, we use the minimum integration time ($t_{\mathrm{min}}$), which is the integration time ($t_{\mathrm{int}}$) to achieve a signal-to-noise ratio (SNR) of one; with the SNR defined as the ratio between the signal and noise power~\cite{Vigneau2023}. We use the $N\leftrightarrow N+1$ Coulomb peak as shown in Fig.~\ref{fig:4}(a). We define the signal level at the top of the peak and the background level away from the peak, representing a charge sensing shift of more than a linewidth. At each of these two points, we obtain time traces of the in-phase (I) and quadrature (Q) signals, which we plot in the I-Q plane as shown in the inset of Fig.~\ref{fig:4}(a). The lowest integration time of the experimental setup (4~ns) is defined by the 250~MHz sampling rate of the analog-to-digital converter. For higher $t_{\mathrm{int}}$, we downsample the time series data using boxcar averaging, with the $t_{\mathrm{int}}$ then given by $t_{\mathrm{int}}\cdot W_{\mathrm{BC}}$ where $W_{\mathrm{BC}}$ is the width of the boxcar used, see Section \ref{sec:Methods:SNR_tmin}. We calculate the SNR using,


\begin{equation}
    \mathrm{SNR} = \frac{\left(I_{\mathrm{on}}-I_{\mathrm{off}}\right)^{2}+\left(Q_{\mathrm{on}}-Q_{\mathrm{off}}\right)^{2}}{0.25\left(\sigma_{\mathrm{on}}+\sigma_{\mathrm{off}}\right)^{2}},  
\label{eq:SNR}
\end{equation}

\noindent where $\sigma_{\mathrm{on,off}}$ are the 2D standard deviations extracted from a Gaussian fit to each of the signal and background blobs in the IQ plane~\cite{von_horstig_multimodule_2024}. 

In Fig.~\ref{fig:4}(b), we present a log-log plot of the SNR as a function of $t_{\mathrm{int}}$ with a linear fit to extract $t_{\mathrm{min}}$ by extrapolating to SNR = 1 for each applied power. The linear dependence indicates that the noise is likely white noise generated by the cryogenic amplifier. We then plot $t_{\mathrm{min}}$ as a function of power in Fig.~\ref{fig:4}(c), where at each power the excitation frequency has been optimized. We find a monotonically decreasing $t_\text{min}$ as the power increases. In the low-power regime, we observe the expected P$^{-1}$ dependence~\cite{Vigneau2023}. However, for powers above -82~dBm at the input of the resonator, $t_\text{min}$ decreases more rapidly with a P$^{-2.5}$ dependence. Considering the frequency versus power plot in Fig.~\ref{fig:4}(d), we see how the resonance dip (and hence $L_{\mathrm{K}}$) changes as a function of rf power. Specifically, we see that the resonator starts to become strongly non-linear at higher powers. In this non-linear regime, as well as the resistive change resulting from the SET Coulomb peak, there is also a non-linear inductance change associated with the large rf current flowing through the inductor. This enhancement in readout sensitivity in the parametric regime has also recently been reported in a system where a Josephson junction array resonator was used ($t_\text{min}=10$~ns)~\cite{havir_near-unity_2025}. At the highest powers measured here, before the superconductor turns normal, we find a lowest $t_{\mathrm{min}}$ value of $1\pm0.3$~ps. In general, the integrated rfSET based on the kinetic inductance of TiN not only demonstrates a reduction of more than four orders of magnitude in the sensor footprint, but also represents an improvement in $t_{\mathrm{min}}$ of more than two orders of magnitude compared to the state-of-the-art of 625~ps~\cite{Keith2019}. We achieve the latter by exploiting a new regime in semiconductor spin qubit readout that utilizes the non-linear inductance of kinetic inductors.


\section{\label{sec:4}Conclusions}

We have demonstrated a compact superinductor realised within a silicon integrated circuit, allowing us to demonstrate a radical improvement both in area and sensitivity of the most widely used quantum sensor for semiconductor spin qubits, the rfSET. The next step will be to integrate the resonator with multi-gate transistors~\cite{Ibberson2024} to demonstrate high-speed high-fidelity spin readout well below the microsecond timescale. However, the applications of the TiN-based CMOS inductors could expand well beyond quantum sensing and computing for spin qubits, impacting areas as varied as radio-astronomy, low-power cryoelectronics and quantum sensing and simulations. 

For radioastronomy, photon detectors based on kinetic inductance are widely used in the far-infrared and X-ray frequency bands. A CMOS-based kinetic inductance detector could enable new compact array designs with increased pixel resolution while reaching single-photon resolution, providing an alternative to the more common CMOS single-photon avalanche diodes (SPADs). For cryogenic electronics, the compact nature of the TiN inductor could replace the common spiral inductors, reducing chip area and hence cost. For example, common low-noise cryogenic amplifiers utilize topologies based on LC resonators which could be replaced by TiN based inductors. In addition, the superconducting nature of the film could provide higher quality factors, ideal for achieving high gain and narrow-band filtering.  For quantum sensing, the availability of non-linear, low-loss inductors on a CMOS process opens up the opportunity to produce quantum-limited parametric amplifiers. Finally, on quantum simulations, the new directions based on 1D and 2D resonator arrays could now be manufactured and produced at scale within an IC.
\\
\\
\indent We believe that achieving these proposed breakthroughs will be facilitated by low-cost manufacturing and ample design software infrastructure provided by the semiconductor industry. 

\section{\label{sec:Methods} Methods}
\subsection{\label{sec:Methods:RF-Reflecto} RF-Reflectometry} 

Readout of the SET in this paper is achieved using rf-reflectometry. The rf signal is supplied using an R\&S SMB100B. This signal then passes to the input port of a directional coupler (Minicircuits ZX30-17-5-S+), with the output port connected to the LO of an IQ demodulator (Analog Devices LTC5584). The couple port is connected to the input line of the dilution refrigerator, which consists of a set of copper-nickel (CuNi) semi-rigid coaxial cables with the signal thermalised at each stage by an attenuator. Specifically there is 10~dB at the Cold Plate, 20~dB at the 4~K stage and 0~dB used at all other stages. At the MXC stage, the input line is connected to the couple port of a directional coupler (Krytar 0.5-8~GHz). The input port of the directional coupler is then connected to the device and the output port to the output line. The output line consists of superconducting (NbTi) semi-rigid coaxial cables with no attenuators between the MXC and 4K stage where the line is connected to a low-noise amplifier (Low Noise Factory LNC0.2\_3B). The remainder of the output line in the fridge then uses copper-nickel (CuNi) semi-rigid coaxial cables. The signal is then amplified at room temperature (Minicircuits ZX60-33LNR-S+ \& ZX60-83LN-S+) before connecting to the rf port of the IQ demodulator. The differential output signals of the IQ demodulator are then connected to a Spectrum M4i analogue-to-digital converter. 

\subsection{\label{sec:Methods:Res_Design}Integrated Resonator Circuit Design}

The integrated resonator circuit was simulated using the industry-standard SPICE simulator Spectre\textsuperscript{TM}. S-parameter analysis was employed to extract the reflection coefficient, $S_{11}$, of the resonator across the operational frequency range (from 0.3~GHz to 1~GHz). The circuit consists of an LC resonator with an embedded SET, AC-coupled to a transmission line feed. The resonator includes a poly-resistor structure acting as an inductor, an MOM capacitor, and a transistor-based SET. An MOM capacitor is also used for line coupling. We model the poly-resistor structure as a non-ideal inductor, with the inductance value determined by the thin-film kinetic inductance. The parasitic elements in the non-ideal inductor model are a series resistor, modelling the contact resistance to the thin-film layer, and a shunt capacitor, representing the coupling between the thin-film and the die substrate. The contact resistance values were extracted from dc measurements of previously fabricated test structures, whereas a range of possible kinetic inductance values was considered based on the geometry of the integrated poly-resistor and using the Mattis-Bardeen formula in the low-frequency limit \cite{tinkham_introduction_2015}. The capacitive coupling to the substrate was extracted from simulations of the foundry-provided poly-resistor model, assuming it remains unaffected by the thin-film superconducting transition. The SET is modelled as a two-state resistor, representing the high and low conductance extremes of the Coulomb blockade regime.  Additional layout-dependent parasitics were extracted and incorporated into the circuit netlist using Calibre xACT parasitic extraction tool. 
The resulting simualtion of the circuit exhibits a change of 3~dB or more in the magnitude of $S_{11}$ between the two resistance states of the SET.

\subsection{\label{sec:Methods:SNR_tmin} SNR \& Minimum Integration Time} 

The Signal-to-Noise Ratio (SNR) is defined as the ratio of the signal power (measured from top to valley of a Coulomb blockade peak) to the power associated with the background noise. The minimum integration time ($t_{\mathrm{min}}$) is then defined by the integration time for which SNR = 1, i.e. it represents the integration time below which we are no longer able to distinguish the signal above the background noise level. To calculate the SNR, time traces are taken at the top of the Coulomb peak and then at the background level away from the peak. This then gives the two Fresnel lollipops in the IQ plane as shown in the inset of Fig.~\ref{fig:4}(a). We use the IQ data for each case to create a 2D histogram which can be fitted using a 2D Gaussian of the form given in Eq.~\ref{eq:2D_Gauss}

\begin{equation}
    Z = A\cdot\mathrm{ exp}\left(\frac{\left(I-I_{0}\right)^{2}}{2\sigma_I^2}+\frac{\left(Q-Q_{0}\right)^{2}}{2\sigma_Q^2}\right).
\label{eq:2D_Gauss}
\end{equation}

The output of the fit can then be used to calculate the SNR as given by Eq.~\ref{eq:SNR} where $I_{\mathrm{peak/background}}$ and $Q_{\mathrm{peak/background}}$ come from the $I_{0}/Q_{0}$ values of the relevant fit and $\sigma_{\mathrm{peak/background}}$ takes the average value of $\sigma_{I}$ and $\sigma_{Q}$. To calculate the minimum integration time ($t_{\mathrm{min}}$) from SNR we need to consider the equivalent noise bandwidth (ENBW) of the measurement. This bandwidth depends on any low-pass filter with cut-off frequency, $f_\text{LP}$,  applied to the measurement path and, if averaging is used, then also the number of averages ($N_{\mathrm{avg}}$) as given in Eq.~\ref{eq:ENBW}, where $\eta$ is a prefactor determined by the order of the LPF being applied, 

\begin{equation}
    \text{ENBW} = \frac{\eta \cdot f_\text{LP}}{N_{\mathrm{avg}}}.  
\label{eq:ENBW}
\end{equation}

We then calculate $t_{\mathrm{int}}$ from,

\begin{equation}
t_{\mathrm{int}} = \frac{1}{2\cdot\mathrm{ENBW}}
\end{equation}

For the case of measurements presented in this paper, the sample rate was 250~MHz, and no LPF was used, giving $t_{\mathrm{int}}$ = 4~ns at $N_\text{avg}=1$. This integration time can be increased by randomizing the data and then applying boxcar averaging with a given window size. Using this method, we obtain the data given in Fig.~\ref{fig:4}(b), showing SNR as a function of integration time. Applying a linear regression to the data, we can then extract $t_{\mathrm{min}}$ by extrapolating down to SNR = 1. 

\section{Acknowledgement}
We thank Charan Kocherlakota, Debargha Dutta, and Jonathan Warren of Quantum Motion for their technical support and thank Nigel Cave at Global Foundries for fruitful discussions. 
T.H.S acknowledges the Engineering and Physical Sciences Research Council (EPSRC) through the Centre for Doctoral Training in Delivering Quantum Technologies [EP/S021582/1]. F.E.v.H. acknowledges funding from the Gates Cambridge fellowship (Grant No. OPP1144) and the Engineering and Physical Sciences Research Council (EPSRC) via the Cambridge NanoDTC (EP/L015978/1). A.G.-S. acknowledges an Industrial Fellowship from the Royal Commission for the Exhibition of 1851. M.F.G.Z. acknowledges a UKRI Future Leaders Fellowship [MR/V023284/1].

\section{Author Contributions}

T. H. S., J. K., G. M. N. and G. A. I. acquired the data. T. H. S., J. K., G. M. N. and G. A. I. analysed the data. M. F. G. Z. conceived the experiment. A. G. S. conceived and designed the superconducting structures. F. O. designed the integrated circuits. A. G. S., J. J. L. M. and M. F. G. Z. supervised the work. All authors contributed to the writing of this manuscript.  

\section{Competing interests}

MFGZ is an inventor of a relevant patent (EP3958188B8).


\bibliography{General}

\end{document}